\documentstyle[11pt]{article}

\def\8{\infty}

\def\ra{\rightarrow}

\def\be{\begin{equation}}
\def\ee{\end{equation}}
\def\bea{\begin{eqnarray} & &}
\def\eea{\end{eqnarray}}

\def\lsim{\mathrel{\rlap{\lower3pt\hbox{\hskip0pt$\sim$}}
    \raise1pt\hbox{$<$}}}         
\def\gsim{\mathrel{\rlap{\lower4pt\hbox{\hskip1pt$\sim$}}
    \raise1pt\hbox{$>$}}}         

\newcommand{\beq}{\begin{equation}}
\newcommand{\eeq}{\end{equation}}

\renewcommand{\Im}{\mbox{Im}\,}

\newcommand{\MeV}{\,\mbox{MeV}}
\newcommand{\matel}[3]{\langle #1|#2|#3\rangle}

\newcommand{\La}{\overline{\Lambda}}
\newcommand{\Lam}{\Lambda_{QCD}}

\newcommand{\re}[1]{ref.~\cite{#1}}
\newcommand{\eq}[1]{eq.\hspace*{.1em}(\ref{#1}) }

\begin{document}
\begin{flushright}
\large{
TPI-MINN-94/4-T\\
UMN-TH-1239/94\\
CERN-TH.7171/94\\
UND-HEP-94-BIG\hspace*{0.1em}03}\\
February, 1994\\
\end{flushright}
\vspace{.4cm}
\begin{center} \LARGE
{\bf The Pole Mass of The Heavy Quark.\\
Perturbation Theory and Beyond}
\end{center}
\vspace*{.4cm}
\begin{center} \Large
{I.I. Bigi $^{a,b}$, M.A. Shifman $^c$,
N.G. Uraltsev $^{a,d}$, A.I. Vainshtein $^{c,e}$}\\
\vspace{.4cm}
{\normalsize\it $^a$TH Division, CERN, CH-1211 Geneva 23,
Switzerland \footnote{During the academic year 1993/94}}\\
{\normalsize\it $^b$ Dept. of Physics,
University of Notre Dame du
Lac, Notre Dame, IN 46556, U.S.A.\footnote{Permanent address}}
\\
{\normalsize\it $^c$
Theoretical Physics Institute, Univ. of Minnesota,
Minneapolis, MN 55455}\\
{\normalsize \it $^d$
St. Petersburg Nuclear Physics Institute,
Gatchina, St. Petersburg 188350, Russia $^2$}
\\
{\normalsize\it $^e$ Budker Institute of Nuclear Physics, Novosibirsk
630090, Russia}\\
\vspace{.4cm}
{\normalsize e-mail addresses:}\\
{\normalsize \it BIGI@CERNVM; SHIFMAN@VX.CIS.UMN.EDU;
VAINSHTE@VX.CIS.UMN.EDU}\\

\end{center}
\thispagestyle{empty} \vspace{.4cm}

\centerline{\Large\bf Abstract}
\vspace{.4cm}

The key quantity of the heavy quark theory is the quark mass
$m_Q$.  Since quarks are unobservable one can suggest different
definitions of $m_Q$. One of the most popular choices is the pole
quark mass routinely used in perturbative calculations and in some
analyses based on heavy quark expansions. We show that
no precise definition of the pole mass
can be given in the full theory once
non-perturbative effects are included. Any definition of this quantity
suffers from an intrinsic
uncertainty of order $\Lam /m_Q$. This fact is succinctly
described by the
existence of an
infrared renormalon generating a factorial divergence in the
high-order coefficients of the $\alpha_s$ series;
the corresponding singularity in the Borel plane is situated at $2\pi /b$.
A peculiar
feature is that this renormalon is not associated with
the matrix element of a local operator.
The difference $\La \equiv M_{H_Q}-m_Q^{pole}$
can still be defined in
Heavy Quark Effective Theory, but only at the
price of introducing an explicit dependence on a
normalization point $\mu$: $\La (\mu )$.
Fortunately
the pole mass $m_Q(0)$ {\em per se} does not  appear in calculable
observable quantities.

\newpage
\addtocounter{footnote}{-2}


\section{Introduction}

Significant progress has been achieved recently  in the
quantitative treatment of the decays of heavy flavor
hadrons by employing
expansions in powers of $1/m_Q$,
where $m_Q$ denotes the heavy quark mass.
Exclusive  transitions between two heavy flavor hadrons
are conveniently dealt with by using
the spin-flavor
symmetry of heavy quarks \cite{IW} (see also \re{SV}); the
formalism of Heavy
Quark Effective Theory (HQET)
\cite{HQET,HQETr} incorporates this symmetry
concisely at the Lagrangian level. Inclusive
decays, on the other hand -- nonleptonic, semileptonic or radiative
transitions to any type of the final state --
can be treated directly in QCD \cite{volo1,chay,BUV,BS,BBSUV,BKSV}
via Wilson's operator product expansion \cite{wilson}.

The key parameter in most aspects of heavy quark physics is
obviously `the heavy
quark mass'. There are no problems in defining this quantity
within purely perturbative calculations. At this level one choice
turns out to be particularly convenient: it is the so-called
pole mass of the heavy quark,
defined as the position of the pole in the quark
propagator in perturbation theory.
This quantity, introduced in QCD in the seventies (see, e.g. Ref.
\cite{novikov}), is well defined in
each {\em finite} order of perturbation theory;
unlike many other definitions, it is
introduced in a gauge invariant way.
This convenient feature has made it very popular.
Important results of
perturbative calculations such as the
total semileptonic widths (including radiative corrections) are
routinely
expressed in terms of the pole mass (see e.g. \cite{alt}).

In this paper we exhibit an important drawback in the
concept of a pole mass that becomes apparent as soon
as one addresses leading non-perturbative corrections to order
$1/m_Q$.
The
problem arises,
of course, because the pole mass is sensitive to large distance
dynamics -- although this fact is not obviously seen in the standard
perturbative
calculations -- and the corresponding  treatment requires special
care. It had actually been noted before that marrying full
QCD with the notion of the pole mass faces subtle difficulties
\cite{bigi}.

Our main assertion will be three-fold:

{\bf (A)}
Infrared contributions lead to an intrinsic
uncertainty in the pole mass of order $\Lambda _{QCD}$,
i.e. an effect of relative weight $1/m_Q$. Perturbation
theory itself produces clear evidence for this non-perturbative correction
to $m_Q^{pole}$. The signal is the peculiar
factorial growth of the high order terms in the $\alpha_s$
expansion
corresponding to a renormalon \cite{renormalon,renormalonr}
residing at
$2\pi /b$ in the Borel plane, where $b$ is the first coefficient in the
Gell-Mann-Low function, $b= (11N_c/3) -(2N_f/3)$.  The
physical reason lying behind
this linear effect is just the ``Coulomb'' energy of the heavy quark.

\noindent A subtle point is that
infrared-renormalon effects can usually be
associated with the expectation values
of some local operators \cite{renormalonr,SVZ} of the corresponding
dimension. This general pattern is not realized in the
case
of the pole mass for the reasons which will be clarified  below.

{\bf (B)} The Heavy Quark Expansion (HQE) yields
for
directly observable quantities such as
the total semi-leptonic width of the heavy flavor hadron decay:
\beq
\Gamma (B\ra  X_u+l+\nu_l )=\frac{G_F^2m_b^5}{192\pi
^3}|V_{ub}|^2
\left[ c\frac{\matel{B}{\bar bb}{B}}{2M_{B}}+
{\cal O}(1/m_b^2) \right]
\label{width}
\eeq
where for definiteness we consider $B$ meson decay and  the
coefficient $c$ is $c=1+{\cal O}(\alpha_s )$.
In the total semi-leptonic
width non-perturbative effects of order $1/m_Q$ are absent (see Refs.
\cite{BUV,BBSUV}), and the corrections start from $1/m_Q^2$
(the factor $(2M_{B})^{-1}$ reflects the relativistic normalization of
the state $|B\rangle$).
While the pole mass is useful
within purely perturbative calculations,
it makes no sense to employ it in such an analysis that aims for an
accuracy
$\sim {\cal O}(\Lam/m_Q)$ or even
$\sim {\cal O}(\Lam ^2/m_Q^2)$ --
simply because it {\em cannot} be unambiguously
defined at order $\Lam/m_Q$.

{\bf (C)}
There is a profound theoretical difference between the pole
mass and the total semileptonic width:
calculation of the latter  can be formulated as an OPE-based
procedure,  ensuring factorization of the large and small distance
contributions;  the pole mass, however,
cannot be treated in such a way.

\noindent On the other hand -- and fortunately --
there is no need to use the pole mass in these inclusive
transition rates. Careful analysis shows that
contrary to popular opinion it is the
running mass $m(\mu )$ with $\mu\gg \Lam$ that naturally enters.
The properly
defined running
mass includes only the effects of momenta
higher than $\mu$; therefore if $\mu$ is chosen sufficiently high
there is no infrared uncertainty in $m(\mu )$. In particular, the
parameter most relevant to the inclusive heavy quark decays
is $m(m_Q)$.  In other problems it may turn out that the running
mass normalized at a different point enters; each particular
case requires its own careful analysis.
It is important however,
that
this normalization point never goes down to a typical hadronic scale
if one
wishes to avoid the corresponding infrared uncertainty in the Wilson
coefficients.
We stress here that the
normalization point
$\mu $ does  {\em not} necessarily coincide with the off-shellness of
the heavy quark inside the heavy hadron.

\noindent
Certainly, the definition of the running mass $m(\mu )$ depends
both on the scheme and the gauge used --  an obvious
inconvenience.  In this respect, however,
the running mass does not differ -- as a matter of principle --
from the running gauge coupling constant $\alpha_s (\mu )$
where an explicit scheme and gauge dependence
first emerges on the third
loop level. The essential difference is
that for $m(\mu )$ this dependence
manifests itself already at the
one-loop level; the observables, however, are independent.

\noindent
Correspondingly, any consistent definition of the parameter $\La$
must explicitly introduce the normalization point $\mu$, so that
$\La$ is actually running, $\La (\mu )$, although the running law
is somewhat unconventional, see eqs. (\ref{mmu}), (\ref{o1a}).

The remainder of this paper will be organized as
follows: in Sect. 2 we restate the
general procedure of separating off the infrared effects within
Wilson's OPE and the relation between the infrared renormalons and
non-perturbative condensates; in Sect. 3 the perturbative series for
the heavy quark mass is discussed;  Sect. 4 is devoted to the infrared
renormalon contribution to the mass; Sect. 5 demonstrates that the pole
mass does not appear in directly observable quantities, such as, say, the
total widths; finally in Sect. 6 we discuss the running of $\La$ and
summarize our results.

\section{Separation of the Infrared Effects within Wilson's OPE}

In this section we remind the reader
of a crucial property of the operator product expansion
\cite{SVZ,NSVZ,shifman}: it provides us with a systematic
separation of infrared and ultraviolet contributions.
This brief
excursion into standard OPE applications will allow us to reveal a
basic difference in how
infrared effects enter in the pole mass and, say, into
correlation functions at large momentum transfer.

As an example let us consider the correlation function of vector
currents,
\beq
\Pi_{\mu\nu} \equiv (q_\mu q_\nu -q^2g_{\mu\nu} )\, \Pi
= i\int d^4 x {\rm e}^{iqx}
\langle 0 |T\{ j_\mu (x) j_\nu (0)\} |0\rangle
\label{pi}
\eeq
where
$$
j_\mu =\bar\psi \gamma_\mu \psi
$$
with $\psi$ being a massless quark field.
To avoid ``external'' logarithms
which are irrelevant for the problem under discussion one usually
deals
with a modified quantity defined as
\beq
\tilde\Pi = -4\pi^2 Q^2(d/dQ^2)\Pi\; , \; \; \;
Q^2= -q^2 .
\label{tildepi}
\eeq
The one- and two-loop graphs determining $\tilde\Pi$ in
perturbation theory are depicted in Fig. 1.  The well-known
calculation of these diagrams yields
\beq
\tilde\Pi = 1+ \frac{\alpha_s(Q^2)}{\pi}.
\label{alpha}
\eeq
The virtual momenta saturating the corresponding loop integrals
are of order $Q$, and if $Q^2$ is chosen to be large the characteristic
virtual momenta are also large. The
result (\ref{alpha}) then represents the short
distance contribution.

The fact that eq. (\ref{alpha}) is correct in perturbative QCD does not
mean,
however, that it is correct in the full theory.  Indeed, in deriving
eq. (\ref{alpha}) one integrates over all gluon momenta $k$,
including
the domain of small $k$ where $k$ is the momentum flowing
through the gluon line in Fig.~1$b$. Of course, this domain gives a
relatively small
contribution to the integral; nevertheless this contribution is
definitely wrong since for small $k$ the Green functions are strongly
modified by non-perturbative effects and have nothing to do with
the perturbative propagators one uses in obtaining eq.~(\ref{alpha}).

The emergence of non-perturbative corrections can
actually be inferred from
perturbation theory {\em per se} if one recalls the presence of the
Landau singularity in the running coupling constant in the infrared
domain (for QCD). This is the essence of the concept
of infrared renormalons
\cite{renormalon,renormalonr}.  The phenomenon is quite simple and
manifests itself in the behavior of the high order
terms in the $\alpha_s$ expansion.
The relevant diagrams are those where a chain of loops
has been inserted into the
gluon Green function (Fig. 2).
This chain is equivalent
to replacing the fixed coupling by the
running coupling constant $\alpha_s (k^2)$ in the
integrand and integrating it over with some weight function; the
latter is
given by the remaining propagators in the diagram. This weight function is such
that the integral over $d^4 k$ converges in the ultraviolet domain.
To illustrate how this works
let us use, following Ref. \cite{zakharov},
 a simplified expression for $\tilde\Pi$,
\beq
{\tilde\Pi}_{renorm} \sim Q^2 \int
dk^2\frac{k^2\alpha_s(k^2)}{(k^2+Q^2)^3} .
\label{renorm}
\eeq
This expression coincides with the original one in the limits
$k^2\ll Q^2$ and $k^2\gg Q^2$  relevant to the infrared and ultraviolet
renormalons, respectively.

The integral in eq.~(\ref{renorm}) is saturated at $k^2\sim Q^2$; to
get the
main contribution one
substitutes $\alpha_s (Q^2)$ for $\alpha_s (k^2)$ and
finds the
standard
two-loop result for $\tilde\Pi$. If, however, we are interested in high
orders in $\alpha_s(Q^2)$ the tails at $k^2\ll Q^2$ and $k^2\gg Q^2$
become important. The contribution of these tails have the form
\beq
{\tilde\Pi}_{IR}
=Q^{-4}\int_0^{Q^2} k^2 dk^2\alpha_s (k^2) ,
\label{IR}
\eeq
and
\beq
{\tilde\Pi}_{UV}
=Q^{2}\int_0^{Q^2} \frac{dk^2}{k^4}\alpha_s (k^2) ,
\label{UV}
\eeq
where the subscripts are self-evident. Substituting the running gauge
coupling
$$
\alpha_s (k^2) = \frac{\alpha_s (Q^2)}{1-(b\alpha_s (Q^2)/4\pi)
\ln (Q^2/k^2)} ,
$$
where $b$ is the first coefficient in the Gell-Mann-Low function, and
expanding in $\alpha_s(Q^2)$ we get  the whole series in
$\alpha_s(Q^2)$.
Namely,
\beq
{\tilde\Pi}_{IR} =
\alpha_s(Q^2)\sum_n \left(\frac{b\alpha_s (Q^2)}{4\pi}\right)^n
\int_0^{Q^2}\frac{dk^2 k^2}{Q^4}\left( \ln\frac{Q^2}{k^2}\right)^n
=\alpha_s(Q^2)\sum_n \left(\frac{b\alpha_s (Q^2)}{4\pi}\right)^n\,
\frac{n!}{2^{n+1}},
\label{n}
\eeq
and
\beq
{\tilde\Pi}_{UV} =
\alpha_s(Q^2)\sum_n \left(-\frac{b\alpha_s (Q^2)}{4\pi}\right)^n
\int_{Q^2}^\8 \frac{dk^2 Q^2}{k^4}\left( \ln\frac{k^2}{Q^2}\right)^n
=\alpha_s(Q^2)\sum_n \left(-\frac{b\alpha_s (Q^2)}{4\pi}\right)^n\,
n! .
\label{UVs}
\eeq
The first expression corresponds to the infrared renormalon while the
second one refers to the ultraviolet renormalon. The factorial growth
of the coefficients is explicit in both cases. There are two differences,
however. The ultraviolet renormalon is represented by the
sign-alternating series, in distinction to ${\tilde\Pi}_{IR}$. The second
difference is the $n$ dependence of the coefficients in front of
$n! (b\alpha_s /4\pi )^{n+1}$. In the infrared renormalon this
coefficient is $2^{-(n+1)}$, compared to 1 in the ultraviolet renormalon.
These differences result in the fact that the positions of the
singularities in the Borel plane (to be denoted $\tilde b$) are at
\beq
{\tilde b}_{IR} =\frac{8\pi}{b} ,\,\,\, {\tilde b}_{UV} =-\frac{4\pi}{b} .
\label{pos}
\eeq
In what follows we will not consider the ultraviolet renormalon, as well as
other sources of the singularities in the Borel plane, for instance,
instanton-antiinstanton pairs (see e.g. \cite{Bal}) producing a
singularity at $\tilde b =4\pi$.

The occurrence of the factorial in ${\tilde\Pi}_{IR}$ is correlated
with the fact that at large $n$ the integral is saturated not at
the large scale
$k^2\sim Q^2$ but, instead, at parametrically low scales $k^2\sim
Q^2/n$.
Therefore when $n$ becomes large one encounters the
Landau
singularity, and -- in an indirect way -- comes to the
realization that in the
infrared
domain the gluon Green function cannot coincide with the
perturbative expression.

The factorial growth of the coefficients points to the fact that one
cannot -- as a matter of principle --
infinitely increase the accuracy of the perturbative approximation
by including
higher and higher terms in the series: starting from some order the
absolute
size of the corrections increases again. The best one can
achieve is to
truncate the
series  at an optimal value of $n$ ensuring the best
possible approximation. It is not difficult to check that the error
introduced by this truncation is of the order of
\beq
\exp\{-\frac{8\pi}{b\alpha_s(Q^2)}\}\sim\frac{\Lam^4}{Q^4}.
\label{error}
\eeq
Since the series is not Borel-summable, it is in principle impossible to
achieve a better accuracy within the framework of perturbation
theory. All one can conclude is that there must be a non-perturbative
effect
of the form (\ref{error}).

It is well known that this problem does not mean that one has to
abandon hope for establishing
theoretical control over the terms of this order of
magnitude. The way out is provided by a
consistent use of Wilson's OPE procedure.
First one introduces a momentum scale $\mu $ such that
momenta above $\mu$ can be treated perturbatively. All
loop momenta are classified according to whether they
exceed $\mu$ or lie below it. The integration over the
infrared domain below $\mu$ is then explicitly
{\em excluded}
from the perturbative calculation and one finds
\beq
{\tilde\Pi}_{pert} \equiv \tilde\Pi |_{k>\mu }
= 1+ \frac{\alpha_s(Q^2)}{\pi} -{\rm
Const}\,\frac{\alpha_s(\mu^2)\mu^4}{Q^4}.
\label{alphamu}
\eeq
The fact that the subtracted term in this particular case is
proportional
to $\mu^4$ is not
accidental, of course: it can be anticipated from
eq. (\ref{error}) and it will be
clarified shortly. This result automatically follows from the explicit
calculation,
provided that the scale $\mu$ is
introduced without breaking the gauge invariance of the theory
(in practice however this may turn out to be a technically
highly non-trivial exercise).
Excluding the domain $k<\mu$ from the perturbative calculation
does not mean that we just lose this contribution. Within the
Wilson
OPE the contribution of this infrared domain re-enters through
the vacuum expectation value
\beq
\Delta\tilde\Pi = -\frac{\pi}{3Q^4}\langle \alpha_s G^2\rangle
\label{deltapi}
\eeq
where $G= G_{\mu\nu}^a$ is the gluon field strength tensor, and the
operator $G^2$ in the right-hand side is normalized at $\mu$ (i.e. by
definition this vacuum expectation value includes all virtual
momenta below $\mu$). The vacuum expectation value
$\langle \alpha_s G^2\rangle$ has the form
\beq
\langle \alpha_s G^2\rangle = {\rm Const}\,\Lam^4 +
{\rm Const}\,\alpha_s(\mu^2)\mu^4
\label{condensate}
\eeq
and in the full expression
${\tilde\Pi}_{pert}+\Delta\tilde\Pi$ the
dependence on the auxiliary parameter $\mu$ cancels.
The requirement that the dependence on $\mu$
cancels in the end dictates that $\mu$ must
appear in ${\tilde\Pi}_{pert}$ as $\mu^4$, since no gauge invariant
operator of lower dimension exists.

Let us emphasize that the normalization point $\mu$
should be high enough to ensure the applicability of perturbative
calculations above $\mu$, i.e.
$$\alpha_s (\mu )\ll 1.
$$
On the other hand,  it is desirable to have this parameter as
small as possible, so that the corresponding terms
will
represent insignificant corrections  in
eqs.  (\ref{alphamu}),
(\ref{condensate}).  If this wish can be satisfied there is no need to
carefully work out details of how to introduce $\mu$ explicitly!
In particular, in eq. (\ref{condensate}) the term with
$\Lam^4$ should be much larger than that with $\mu^4$.
In other words, the numerical values of the condensates are assumed
to be much larger than their perturbative parts for some $\mu$
belonging to the window discussed above.
This
is what is called the practical version of OPE -- powers of $\mu$ do
not show up explicitly then, although, of course, the conceptual
necessity of having $\mu$ should be always kept in mind ($\mu$ is
certainly  kept in logarithms in the practical version of OPE). It is
fortunate for applications of  QCD that the practical version of
OPE works well in the
vast majority of instances; this could not have been anticipated {\em a
priori}.
Otherwise,  all calculations based on OPE
would be much less useful since it would be mandatory to explicitly
construct the procedure of introducing $\mu$.

What then happens with the infrared renormalon within Wilson's
procedure?  If one defines the perturbative part with the infrared
cut off at the point $\mu$, the factorial growth of the coefficients in
the
perturbative series in $\alpha_s$ stops starting from some value
of $n$, $n \sim Q^2/\mu^2$, since the integral ``wants''
to be saturated in the domain $k^2 < \mu^2$ and this
domain is now simply eliminated from the perturbative sector. The price
one has to pay is the introduction of a new, non-perturbative
parameter,
the gluon
condensate. Once it is introduced, however, the perturbation theory
is amended, the effects $\propto Q^{-4}$ become tractable and this
accuracy is legitimate. If necessary, one goes a step further. Of
course, corrections of higher order in $Q^{-2}$ require the
introduction of new higher-dimensional
condensates.

Let us emphasize that the actual {\em uncertainty} due to the
infrared renormalon, eq.~(\ref{error}),
should by no means be equated with the contribution from the
gluon condensate. It is true that they are of the same order
in the parameter
$\Lam /Q$ and the infrared renormalon anticipates the appearance of
the gluon
condensate; the gluon condensate contribution, however, is much
larger {\em numerically}, and this is the reason why the practical version
of OPE is so successful. One can interpret this fact in terms of
extremely strong distortions of the Green functions in the infrared
domain \cite{SVZ}.  In other words the modifications (compared to
smoothly
extrapolated perturbative Green functions) are not just of the order
of unity
but are much larger {\em numerically}.

The discussion above makes it clear that
in principle perturbation theory already signals the emergence of a
non-perturbative correction in $\tilde\Pi$ of order $\Lam^4/Q^4$.
A hint is provided by the
simplest diagram of Fig. 1$b$. If we introduce $\mu$ and
calculate the graph of Fig. 1$b$ according to the Wilson procedure we
discover a
correction $\alpha_s\mu^4$.  In combination with the general fact of
strong
distortion of the Green functions in the infrared domain this
$\alpha_s\mu^4$ correction triggers the rest of the machinery which
eventually leads to the
non-perturbative gluon condensate contribution in $\tilde\Pi$.

The lessons one can draw from this rather standard procedure
\cite{SVZ,NSVZ,shifman}
are
quite evident: although it is impossible to actually calculate
non-perturbative contributions by analyzing the behavior of
perturbation theory for a particular quantity, the fact that
non-perturbative contributions exist and their particular form  can be
inferred. Moreover, one can find out what kind of non-perturbative
terms are
to be expected in a two-fold way: by studying low order
perturbative graphs within the Wilson procedure and by inspecting
high orders of
perturbation theory as they are generated by the infrared
renormalon.

Below we will use the both lines of reasoning  to show that
the ``pole mass'' of heavy quarks  -- if treated in the
context of problems where
we intend to include
non-perturbative effects --  contains a piece of the order of
$\Lambda_{QCD}$ which must be
considered as an intrinsic uncertainty. Unlike the standard case,
however, this piece is {\em not} related to any matrix element of a
local gauge invariant operator. Therefore, one cannot
amend the perturbation theory based on
$m_Q^{pole}$ in the manner it is usually
done. This happens due to the fact that the notion of the
``pole mass''  by itself is ill-defined.

\section{Perturbative Corrections to $m_Q$}

Following the general strategy outlined above we start our
analysis of the pole mass with the simplest perturbative graph of
Fig. 3. Instead of a straightforward calculation of this graph~\footnote{The
corresponding computation can, of course, be found in any textbook
on quantum electrodynamics (QED); to this order there is no
difference between QED and QCD, up to a trivial overall
factor.} we follow the Wilson procedure and introduce, first of
all, the
normalization point $\mu$ such that the domain of virtual
momenta $k<\mu$ is discarded. Thus, we are going to calculate
an analog of ${\tilde\Pi}_{pert}$ in eq. (\ref{alphamu}).

As already mentioned, separating the infrared and ultraviolet
domains would in general require a
rather sophisticated machinery~\footnote{As will be seen below,
naive
dimensional regularization cannot serve this purpose.}.
Fortunately,
the situation simplifies in this particular case
because the diagram of Fig. 3 is the same as
in QED. In QED we can just introduce a ``photon mass'' $\lambda$,
which, on the one hand, preserves Ward identities associated with
current conservation, and on the
other, suppresses the contribution of all virtual momenta below
$\lambda$. In this way, the infrared domain is automatically
discarded. The photon mass $\lambda$ is to be identified with the
normalization point $\mu$, cf. eq. (\ref{alphamu}).  The
procedure of introducing $\mu$ suggested here cannot be extended to
higher loops. It is quite satisfactory, however, for our
more limited
purpose, namely to establish the presence of a
correction of order $1/m_Q$
in the pole mass.

Accordingly we use the following expression for
the gluon propagator
\beq
D^{ab}_{\mu\nu}(k)=-\delta^{ab}
\left( g_{\mu\nu} -\xi \frac{k_\mu k_\nu}{k^2}\right)
\frac{1}{k^2-\lambda^2}\frac{M_0^2}{M_0^2-k^2}.
\label{gluon}
\eeq
Here $\xi$ is the gauge fixing parameter and the term
$M_0^2/(M_0^2-k^2)$ ensures ultraviolet regularization ($M_0$
is the ultraviolet cut-off).
We have introduced it to be safe in the ultraviolet
regime, although it will
play no role in what follows. Since we are interested in the infrared
domain we further assume that
$$
\frac{\lambda}{m_Q} \ll 1 .
$$

Explicit evaluation of the diagram of Fig. 3 yields
\beq
m_Q(\lambda) = m_Q^{(0)}\left(
1+\frac{\alpha_s}{4\pi}(\ln\frac{M_0^2}{{m_Q^{(0)}}^2}
+{\rm Const.})\right) -\frac{2\pi}{3}\frac{\alpha_s}{\pi}\lambda .
\label{mlambda}
\eeq
Here $m_Q^{(0)}$ is the bare mass (the mass at $M_0$), while the
quantity on the left-hand side is the {\em pole} mass which
depends on
$\lambda$. We remind the reader that the domain
of virtual momenta $k<\lambda$ is absent in $m_Q(\lambda )$, and
this leads
to the dependence on $\lambda$.

Next, we identify $\lambda$ and $\mu$, as explained above, and
express the mass at one normalization point (we will refer to this
mass as the running mass) in terms of that corresponding to some
``starting'' normalization point $\mu _0$,
\beq
m_Q(\mu ) = m_Q (\mu_0) +\frac{2\pi}{3}\frac{\alpha_s}{\pi}
(\mu_0 -\mu ).
\label{mmu}
\eeq
What is important in eq. (\ref{mmu}) is the occurrence of a
correction of order $1/m_Q$ relative to the
leading term.

The procedure outlined above is not  unambiguous in both
its elements -- the definition of the running mass and the specific
manner in which the normalization point has been introduced.
In principle,
one can
use other prescriptions. Let us mention, for instance,  the
suggestion of  Ref. \cite{voloshin} where the running mass $m_Q(\mu
)$ was introduced through a certain integral over the cross section
for the process $\gamma +Q \ra g +Q$. The normalization point $\mu$
then enters as an upper limit in this integral (see
\cite{voloshin} for more details; we plan to discuss this approach in a
forthcoming publication). Within this procedure one gets an
analog of eq.
(\ref{mmu}) which still contains a linear correction,
albeit with a
different numerical coefficient.  While the
numerical value of the $1/m_Q$ correction
is thus scheme-dependent, its presence is not!

The physical meaning of the $1/m_Q$ contribution
is quite transparent. It is
nothing else than the classical Coulomb self-energy of
a static color
source, see e.g. Ref. \cite{bigi}.
The energy of the Coulomb field at a
distance $r_0$ is given by
\beq
E_{Coul} = \frac{2\alpha_s}{3}\frac{1}{r_0}.
\label{coulomb}
\eeq
Identifying $r_0$ with $1/\mu$ we recover eq. (\ref{mmu}).

The Coulomb contribution to the mass can
readily be derived directly from the graph in Fig. 3.
In the limit of large $m_Q$ (compared to the gluon virtual
momentum
$k$), i.e. in the static limit, the expression for the mass shift
takes the form
\beq
\delta m_Q\equiv m_Q^{pole} - m_Q (\mu )=
-i \frac{4}{3} g_s^2\int \frac{d^4k}{(2\pi )^4}
\frac{1}{(k_0 +i\epsilon )(k^2 +i\epsilon )}
\label{delta}
\eeq
with an ultraviolet cut-off in the $k$  integration
implied at $\mu$. (The integral for $\delta m_Q$ becomes linearly
divergent in the static limit.) It is easily seen that
the only surviving contribution in eq. (\ref{delta})
comes from $1/(k_0 +i\epsilon )$ in the form of a term
$-i\pi\delta (k_0)$ . The
fact that
the static limit implies the vanishing of $k_0$ is quite evident
by itself; the occurrence of the linear infrared effect under
consideration can be traced back to this feature of the static
limit.

Performing first the integration over $k_0$ (which reduces
merely to
putting $k_0 =0$) we arrive at
\beq
\delta m_Q = \frac{8\pi}{3}\alpha_s \int
\frac{d^3k}{(2\pi)^3}\frac{1}{{\vec k}^2}.
\label{deltak}
\eeq
If an ultraviolet cut-off is introduced through a factor
$\mu_0^2/(\mu_0^2+{\vec k}^2)$ in the integrand we get
\beq
\delta m_Q = m_Q^{pole} - m_Q (\mu_0) = \frac{2}{3}\alpha_s\mu_0
,
\label{deltamu}
\eeq
cf. eq. (\ref{mmu}) where, to get the pole mass on the left-hand
side,  one should set $\mu =0$.
Were the cut-off introduced in a ``hard'' way we would get a
different
coefficient in front of the linear term, of course, but the very
fact of its presence would remain intact.

The appearance of the term linear in $\mu$ in eq.~(\ref{mmu}) tells
one, according to the discussion of Sect. 2, that there is a renormalon
singularity in the perturbative series giving rise to a relative
uncertainty in the pole mass of order $1/m_Q$. Below we will
demonstrate it explicitly.

To conclude this section an important remark of a conceptual
nature is in order. The calculation carried out above clearly reveals
an important fact: while the OPE-like procedure
routinely used in HQET
resembles closely Wilson's OPE procedure \cite{wilson},
it has one very
distinct feature. In the static limit all {\em energy} transfers
to the heavy quark line vanish, implying that the time separations
are always large. Physically that is quite transparent: the heavy quark
after being placed at the origin as a static color source at
$t=-\infty$ remains there at rest till $t=+\infty$. The OPE
procedure in the effective low-energy theory is then based on a
separation of  small
{\em spatial} from large {\em spatial} momenta. Therefore,
below $m_Q$ we actually deal with a three-dimensional version of
OPE, which results in peculiarities that might seem strange, at first
sight, to those who got used to the standard features of the
four-dimensional OPE.

\section{Infrared Renormalon}

Next one examines the impact of the
high order corrections in $\alpha_s$ to the
pole mass. We again study the chain of loops inserted
into the gluon propagator,
this time in the graph of Fig. 3, see Fig. 4. Summing all these
``bubbles''
amounts to replacing $\alpha_s$ in eq. (\ref{deltak})
by the running coupling $\alpha_s ({\vec k}^2)$ in the
integrand,
\beq
m_Q^{pole} - m_Q(\mu_0) =
\frac{8\pi}{3}\int_{|{\vec k}|<\mu_0}
\frac{d^3k}{(2\pi)^3}\frac{\alpha_s({\vec k}^2)}{{\vec k}^2} .
\label{mpoler}
\eeq
The running gauge coupling is given by
\beq
\alpha_s ({\vec k}^2) =
\frac{\alpha_s (\mu_0^2)}{1-(b\alpha_s(\mu_0^2)/4\pi )
\ln (\mu_0^2/{\vec k}^2)} =
\alpha_s (\mu_0^2)\sum _{n=0}^{n=\infty}\left(
\frac{b\alpha_s (\mu_0^2)}{4\pi}\ln\frac{\mu_0^2}{{\vec
k}^2}\right)^n .
\label{alphamu0}
\eeq
Substituting the expansion of eq. (\ref{alphamu0}) into eq.
(\ref{mpoler})
we immediately obtain the following series:
\beq
m_Q^{pole} -m_Q (\mu_0) =
\frac{4\alpha_s (\mu_0)}{3\pi}\mu_0
\sum C_n\left(\frac{b\alpha_s (\mu_0 )}{4\pi}\right)^n
\label{mpoleren}
\eeq
where the coefficients $C_n$ are given by
\beq
C_n =\int_0^1 dx \left( \ln\frac{1}{x^2}\right)^n .
\label{cn}
\eeq
At large $n$ these coefficients grow factorially,
\beq
C_n = 2^n n! \;\;.
\label{cnlarge}
\eeq
In other words one can say that the position of the nearest
singularity in the Borel plane \cite{renormalon,renormalonr}
is at $\tilde b = 2\pi /b$.
This series is not Borel-summable due to the presence
of an infrared renormalon;
truncating it at the optimal value of $n_0$
\beq
n_0\sim \frac{2\pi}{b \alpha_s(\mu_0^2)}
\label{nopt}
\eeq
one arrives at an estimate of the irreducible
uncertainty in $m_Q^{pole}-m_Q (\mu_0)$,
\beq
\Delta (m_Q^{pole}-m_Q (\mu_0)) \sim \frac{8}{3b}\mu_0
\exp\{- \frac{2\pi}{b\alpha_s (\mu_0 )}\} \sim \frac{8}{3b} \Lam .
\label{linear}
\eeq

Thus, we see that perturbative QCD indeed does not allow one to
define
$m_Q^{pole}$ with an accuracy better than $\Lam$,
i.e. an infrared effect
linear
in $\Lam/m_Q$.

A simple way to explain the estimate (\ref{linear}) is to turn to the
original integral (\ref{mpoler})
\beq
m_Q^{pole}-m_Q (\mu_0) =
\frac{4}{3\pi}\alpha_s (\mu_0)
\int_0^{\mu_0}
\frac{dk}{1 -\frac{b\alpha_s (\mu_0)}{4\pi} \ln \frac{\mu_0^2}{k^2}} .
\label{mpolerb}
\eeq
Previously we have just expanded the denominator in the powers of
$\alpha_s (\mu_0)$, obtaining in this way the factorial behavior
of the expansion coefficients related to the existence of the pole
singularity in the integrand. Now, instead, we regularize the
singularity, say, by adding $i\epsilon$ in the denominator. Then eq.
(\ref{mpolerb}) acquires the imaginary part,
\beq
\Im (m_Q^{pole}-m_Q (\mu_0)) = \frac{8\pi}{3b}\Lam
\label{IM}
\eeq
where $\Lam$ parametrizes the position of the infrared pole in the
running gauge coupling. The estimate (\ref{linear}) above coincides with
eq. (\ref{IM}), up to a factor $1/\pi$ reflecting the difference between
the real and imaginary parts.

(Let us parenthetically note that in eq. (\ref{mpoler}) we substituted the
soft cut-off $\mu_0^2/(\mu_0^2 +{\vec k}^2)$ by a step function at
$|\vec k | =\mu_0$. This is unimportant for the infrared renormalon where
the integral is saturated at  ${\vec k}^2  \sim\mu_0^2 /n$. However, with the
soft cut-off restored the very same integral (\ref{mpoler})
produces the ultraviolet renormalon due to the domain
${\vec k}^2 \sim\mu_0^2 n$. It is not difficult to check that the
corresponding factorial behavior is the same, up to a sign,
$(C_n)_{UV} = (-2)^n n!$. In contrast to the situation with the
polarization operator $\tilde\Pi$ considered in Sect. 2, these two
singularities in the Borel plane, infrared and ultraviolet, are symmetric
with respect to the origin.)

The statement above -- the impossibility of defining
$m_Q^{pole}$ to the accuracy better than
$\Lam$ -- implies some  tacit
assumptions. In
particular, we assumed that one should use the value of the running
coupling
$\alpha_s({\vec k}^2)$ in the integrand in \eq{mpoler}. This natural
prescription is easily justified in QED where the Ward identity
reduces
the renormalization of the coupling constant to the corrections to
the photon  propagator.  In non-abelian theories this is not the case
in covariant
gauges.  A general argument below illustrates
the fact that one cannot get  anything else.

Being interested in effects occurring at the scale $\mu$, much
below the mass
of the heavy quark, one integrates out all momenta above $\sim
\mu$ and arrives at an effective field theory of a nonrelativistic
heavy quark with an ultraviolet
cutoff $\mu$. The parameters of
QCD are then
$\alpha_s(\mu)$ and $m_Q(\mu)$ (and, in principle,
masses for the ``light'' quarks);
including external interactions adds also
the
corresponding couplings that must be renormalized at the  same
scale
$\mu$ as well. The dependence of these parameters defining
the  effective theory on
the renormalization point $\mu$ follows from the requirement
that physical observables do
{\em not} depend on $\mu$.
If the renormalization
procedure is such  that lowering $\mu$ from a value $\mu_1$
down to
$\mu_2$ incorporates radiative corrections due to
virtual momenta $|\vec k|$ between $\mu_1$
and
$\mu_2$, then there must be a linear dependence of $m_Q(\mu)$ on
$\mu$, necessarily of the form
\beq
\frac{dm_Q(\mu)}{d\mu}= \beta_m(\alpha_s (\mu )) =-\beta_m^{(1)}
\alpha_s(\mu) + ... \; .
\label{o22}
\eeq
The fact that it originates from radiative corrections  is indicated  by
the
explicit factor
$\alpha_s$ on the right-hand side. Of course, the exact form of the function
$\beta_m(\alpha_s)$
depends on the particular renormalization scheme. If one uses a
scheme that coincides to one loop with the prescription of
introducing a ``gluon mass'' $\mu$, then according to
\eq{mlambda}
\beq
\beta_m^{(1)}= \frac{2}{3} .
\label{o23}
\eeq
The renormalization group equation (\ref{o22}) combined with
Eq. (\ref{o23})
is equivalent to Eq. (\ref{mpoler}): ${\cal O}(\alpha_s^2)$ terms in the
function $\beta_m$
(\ref{o22}) are neglected as subleading.

Recalling the discussion in Sect. 2  one is tempted
to relate the infrared part in $m_Q^{pole}$ in eq.~(\ref{linear})
to the matrix element of some local gauge invariant operator. Alas,
such an  attempt is doomed to fail.
This is most easily seen
by inspecting the local operators of the relevant dimension.
The only potential candidate is
$$
\bar Q iD_0 Q
$$
where $iD_0 = i\partial_0 +g_s A_0$ ( we imply gauge invariance
plus a static description of the field $Q$ similar
to that used in HQET). However, the equations of motion reduce this
operator to those of dimension 5 which can generate corrections
of the relative weight $\Lam^2/m_Q^2$ only,
rather than  $\Lam /m_Q$,
provided that the definition of the quark mass is properly adjusted,
so that there is no so-called residual mass \cite{N51}, see below.

{}From the derivation given above (see Sect. 3 and, especially, the
fact that $k_0 = 0$, as emphasized there) it is clear why
the standard OPE program is inapplicable to
$m_Q^{pole}$. By analyzing Fig. 3 we have realized that only
very small frequencies (of the order of $\mu_0^2/m_Q$) contribute
to the pole mass. In other words, even though the characteristic
{\em spatial} distances are small in the problem at hand, the {\em time}
separation is parametrically
large! To state it in more physical terms:  measuring the pole mass of the
heavy
quark requires a  very
long time, inversely proportional to the allowed uncertainty in the
absolute value of the mass.

This means that the non-perturbative infrared contribution
in $m_Q^{pole}$ cannot be expressed  in terms of
a {\em local} condensate, but,
rather,  through a non-local expectation
value. We have encountered with a similar situation previously
\cite{motion} in connection with the so-called temporal distribution
function defined through the hadronic matrix element of the
operator
\beq
(D_iQ(t,\vec x =0)){\rm e}^{-i\int_{0}^t A_0 (\tau )d\tau }
(D_iQ(t=0,\vec x =0))
\label{string}
\eeq
where $Q$ is the heavy quark field and $D_i$ is the spatial component of
the covariant derivative. The pole mass formally appears in consideration
of the operator (\ref{string}) if one considers the matrix element
of this operator over the heavy quark state in the limit $t\ra\8$. Thus,
we deal here with a generalization of the standard Wilson path operator
for an open path along the $t$ direction.

\section{Irrelevance of the Pole Mass}

The only systematic approach presently known that allows
one to treat
non-perturbative effects in QCD  analytically  is based on  Wilson's
OPE. This procedure requires -- as discussed in detail in Sect. 2 --
a careful separation of contributions from large and short
distance dynamics. Our findings from
the previous section
suggest that the pole mass cannot be handled within such an approach.
In this section we will demonstrate  that this is indeed the case.

To be specific let us discuss the problem of charmless
semileptonic decays of $B$ mesons
which has been already mentioned in Sect. 1.
In the parton model the total width is given by the probability of
the
$b\ra u l{\bar\nu}_l$ transition; neglecting all corrections we have
\beq
\Gamma_0 =\frac{G_F^2m_b^5}{192\pi^3}|V_{ub}|^2 .
\label{gamma0}
\eeq
When $\alpha_s$ corrections are included the result of the
explicit perturbative computation is most naturally expressed
in terms of the
pole mass (and this is what is usually done, see e.g. \re{alt}),
\beq
\Gamma_{pert} =\frac{G_F^2(m_b^{pole})^5}{192\pi^3}|V_{ub}|^2
\left( 1-\frac{2\alpha_s(m_b^2)}{3\pi}(\pi^2-\frac{25}{4}) +...\right)
{}.
\label{gammapert}
\eeq
This formula is perfectly legitimate as long as we stay in
perturbation theory.  If, however, we would like to take into account
non-perturbative (infrared) effects the use of eq. (\ref{gammapert})
can be seriously misleading. Let us clarify this point.

As explained above, the perturbative series for
$m_b^{pole}$ diverges factorially -- a fact signalling the presence of
a non-perturbative contribution. A similar factorial divergence takes
place in the $\alpha_s$ expansion in the brackets of eq.
(\ref{gammapert}). This perturbative series does have
the renormalon singularity at $\tilde b = 2\pi /b$ giving rise
to (uncontrollable) corrections of order $\Lam /m_Q$. Both effects
combine, however,
to cancel each other, if the running mass $m_Q(m_Q)$ is used.

To see how this cancellation works it is sufficient to consider
one-gluon exchange graphs, Fig. 5. One must keep in mind
that the gluon Green function is assumed to be dressed as in
Fig. 4,
so that $\alpha_s (k^2) /k^2 $ must be used for the gluon
propagator. This gives rise to
the usual infrared renormalon.

We remind the reader that we are interested in the imaginary parts
of the diagrams in Fig. 5 corresponding to
the appropriate cuts (i.e. the
total semileptonic width). The graphs 5$a$ and $b$ contain the effect
of the radiative shift of the mass discussed in Sect. 3, and, in
particular, the factorial divergence, see Sect. 4. Our point is that the
integrands in the Feynman integrals
for the diagrams 5$a$ and $d$ will completely compensate
each other in the domain of virtual  momenta $|k|\ll m_b$;
likewise with the graphs 5$b$ and $e$. The easiest way
to demonstrate this compensation is to contract the fermion loop
in diagram 5$a$ to a local effective vertex $\not\!\!
p^5$ where $p$ is the momentum carried by the heavy quark line.
The net effect of the diagrams 5$a$ and 5$b$
is then to convert the bare $b$ quark mass into
$(m_b^{pole})^5$ in the total width where $m_b^{pole}$ in this
approximation is
\beq
m_b^{pole} = m_b^{(0)} -\Sigma
\label{sigma}
\eeq
and $-\Sigma$ is given by the graph of Fig. 3. Explicitly these graphs
produce
\beq
\Delta\Gamma_{a+b}=\frac{G_F^2|V_{ub}|^2}{192\pi^3}
\left( -
5(m_b^{(0)})^4 \Sigma \right) .
\label{DG}
\eeq
The result for $\Sigma$ is implicitly given in eq. (\ref{mlambda});
its explicit expression
is not needed for our purposes here, where we have to consider only
the integrand of the integral determining $\Sigma$.

Let us now turn to the diagrams 5$d$ and 5$e$. First we observe that
the $\not\!\!  p^5$ vertex we had as an effective vertex representing
the light fermion loop in Fig. 5a is actually a
$$
\bar b (i\not\!\!\partial )^5 b
$$
term in the effective Lagrangian. The gauge invariance of the theory
implies that the emission of the soft (i.e. $|k|\ll m_b$) gluon from this
fermion loop -- Figs. 5 $d$ and $e$ -- is completely
taken care of by the substitution
$$
\bar b (i\not\!\!\partial )^5 b \ra \bar b (i\not\!\! D )^5 b
$$
where $D_\mu$ is the covariant derivative. Diagrammatically these
effective vertices are presented in Fig. 6. The gluon emission vertex
in Fig. 6$b$ is
$$
5 \not\!\!  p^4 \not\!\!  A \ra 5(m_b^{(0)})^4 \not\!\!  A .
$$
After these remarks it is straightforward to see that the diagrams
5$d$ and 5$e$ yield
\beq
\Delta\Gamma_{d+e}= \frac{G_F^2|V_{ub}|^2}{192\pi^3}
5(m_b^{(0)})^4 \Sigma.
\label{DGprim}
\eeq
We conclude that the  uncertain contribution
from the domain
$|k|\ll m_b$
present in individual graphs and responsible for the factorial
growth of the coefficients (see Sect. 4) is absent in the sum, eqs.
(\ref{DG}) and (\ref{DGprim}).
Of course, the cancellation described above takes place only to
leading order in $k/m_b$. The residual difference between
the graphs 5$a$ and $d$  shows up at the level of integrals of the
type
$$
\int d^3 k /|\vec k |
$$
which are harmless from the point of view of the linear in $1/m_b$
effect we focus on here.
Let us also note in passing that
the diagrams 5$c$ and 5$f$ do not produce contributions
of  this type and this type of divergence.

Thus, in applying heavy quark theory one has to
avoid the pole mass altogether
and to use, instead, the running mass which naturally appears in OPE;
for only in this case one may hope
to get consistent and well-defined expansions in powers of $1/m_Q$.
The leading operator in the expansion in the
problem at hand is $\bar b (i\not\!\! D)^5 b$. From the consideration
above it follows that the normalization point $\mu = m_b$
is the most natural choice: by adopting this normalization
point we avoid any large logarithms as well as the problem of a
factorial divergence of the type discussed in Sect. 4.
Non-perturbative effects enter through the matrix element of this
operator; they are also represented by matrix elements of other
(subleading) operators, for instance, $\bar b (i\not\!\! D)^3i\sigma G
b$.

Using the
equations of motion one reduces the leading  operator to $m_b^5\bar b
b$,
where both $m_b$ and $\bar b b$ are taken at  $\mu = m_b$. We
then evolve $\bar b b $ down to a low normalization point,
$\mu\ll m_b$; the net effect of this evolution is reflected in a factor
of the type $c(\mu, m_b) =1+ a_1(m_b)\alpha_s (m_b) + a_2
(m_b)\alpha_s^2(m_b) + .... $
which is, anyway, included in the perturbative calculation. This
factor contains also  terms of order $(\mu /m_b)^n$
due to the exclusion of the domain below $\mu$ from the
perturbative calculation. It is important that the power $n$ starts
from $n=2$. These small terms, $(\mu /m_b)^n$,
will contain the series in $\alpha_s (\mu )$, not $\alpha_s (m_b)$.  The
additional $\mu$ dependence occurring in this way will be canceled
by that coming from appropriate local operators, e.g.
$\bar b (i\not\!\! D)^3i\sigma G b$.

 Once the operator $\bar b b$ is evolved
down to a low normalization point we can use the relation
\beq
\bar b b = v_\mu\bar{b} \gamma_\mu b + \frac{1}{4m_b^2}\bar{b}
i\sigma G b +
\frac{1}{2m_b^2}\bar{b}((iD)^2-(ivD)^2) b + {\cal O}(1/m_b^4)
\label{o30}
\eeq
explicitly demonstrating the absence of $1/m_Q$ corrections.
Eq. (\ref{o30})  is valid up to terms representing total derivatives.

In other words, one  integrates out the
momenta
above the scale $\mu$ to get a generic operator product
expansion for the width
\beq
\hat{\Gamma}= c(\mu,m_b)\cdot m_b^5\bar{b}b|_\mu +
c_2(\mu,m_b)\cdot m_Q^3\bar{b} i\sigma G b|_\mu +\ldots ;
\label{o31}
\eeq
the matrix element is taken then by using eq. (\ref{o30}).

A question which immediately comes to one's mind is as follows.
What happens if we first evolve the operator $\bar b (i\not\!\!
D)^5b$ down to a low normalization point and only {\em then} use the
equations of motion. At first sight we would get $m_b^5 (\mu)$, not
$m_b^5 (m_b)$ in this case, and in the limit $\mu\ra 0$ we would
recover $(m_b^{pole})^5$. The loophole in this argument is rather
obvious: the operator $\bar b (i\not\!\! D)^5b$
mixes under renormalization with the operators $\bar b
(i\not\!\! D)^4b$, etc., and accounting for this mixing returns us
to the mass normalized at $m_b$.

It is instructive to dwell on  this issue of mixing in more detail, the
more so since it is intimately related to the notion of the residual
mass, to be discussed below.
Consider the operator $\bar b i\not\!\! Db|_{\mu_0}$ and evolve it
down to $\mu$. The effect of the evolution is described by the
diagrams   of Fig. 6$b$.  If only terms linear in $\mu$ are kept,
the
only relevant graph is the vertex renormalization; this graph is
the same as that for the mass renormalization (Fig. 3), up to a sign.
As a result we get
\beq
\bar b i\not\!\! Db|_{\mu_0} =\bar b i\not\!\! Db|_{\mu}
-\frac{2\pi}{3}\frac{\alpha_s}{\pi}(\mu_0 - \mu )\bar b b.
\label{bDb}
\eeq
Using now the equations of motion in combination with eq.
(\ref{mmu}) we see that both the left- and the right-hand
sides contain $m_b (\mu_0)$, {\em q.e.d.}
A similar relation holds, of course, for any power of $i\not\!\! D$.

Above we have demonstrated our assertions considering explicitly
a certain class of diagrams.  The result is more general, of course. It
can be viewed
as a statement that the inclusive widths of heavy flavors
do
belong to the class of observables which are given by operator
product expansions in
the standard understanding: physics of short and large
distances~\footnote{It
must be noted that it is indeed a statement of separation of large and
short
distances in the process, not of perturbative versus. non-perturbative
effects;
for the Wilson coefficients in general include non-perturbative
contributions
generated for example by instantons of small size $\sim 1/\mu$
or $1/m_Q$.}
can be separated into operators and their coefficients, and the
infrared
behavior of the latter is governed, in turn, by the corresponding local
operators.
The existence of the
$1/m_Q$ renormalon in the heavy quark mass shows that
this statement is not as trivial as it might seem at first sight.
However, as soon  as the
validity of the OPE is accepted, one necessarily arrives at the
irrelevance of the pole mass.

Within the OPE-based approach one obtains the corresponding
inclusive width
as a series of operators of the form $\bar{b}...b$ initially normalized
at
the scale $m_b$; the infrared stability of the inclusive widths
ensures that
the coefficient functions are finite in any perturbative order. Some of
these
operators  contain covariant derivatives acting on the $b$
fields;
due to the equations of motions these derivatives reduce to powers
of the
{\em high} scale mass $m_b(m_b)$. Subleading operators can
produce
relative effects not larger than $1/m_b^2$.

Perturbative corrections actually drop out altogether
from certain observable quantities. Differences in
the lifetimes of different species of hadrons in the
same heavy flavor family provide a prominent example:
the widths of pseudoscalar mesons $P_Q$ and baryons
$\Lambda _Q$ agree through order $1/m_Q$
$$
\frac{\Gamma_{\Lambda _Q}-
\Gamma_{P_Q}}{\Gamma_{\Lambda _Q}+\Gamma_{P_Q}}\sim 1/m_Q^2 ,
$$
in clear  contrast to their masses,
$$
\frac{M_{\Lambda _Q}-M_{P_Q}}{M_{\Lambda _Q}+M_{P_Q}}\sim 1/m_Q.
$$
We plan to  present a more detailed discussion in a forthcoming
paper  \cite{irnew}.

Since
our conclusions obviously do not depend on the particular decay process,
they apply
directly and equally to radiative and nonleptonic decays.

\section{The Running of $\bar\Lambda$ and other Conclusions}

We have discussed in some detail the problem of the pole
mass in
the heavy quark theory.  Due to the peculiarities of the static limit
an infrared term
linear in $\Lam /m_Q$ is generated   in the
pole mass, as signalled  by an infrared renormalon. The
presence of this non-perturbative term makes the notion of the pole
mass,  beyond perturbation theory, not only useless  but, rather,
detrimental. What is even worse, this non-perturbative term cannot
be absorbed into any local condensate, unlike the usual OPE-based
prescription where the
infrared effects are naturally incorporated through
the condensates. Thus, the pole mass should be avoided altogether in
analyzing calculable observable quantities. The problem disappears
provided one uses the running mass $m_Q(\mu)$ normalized at a
sufficiently high point. The same situation in a different context
has been noted recently in Ref. \cite{smith}.

The occurrence of a new, linear infrared effect in the static limit
reminds us of high-temperature QCD. In this theory an external
parameter, temperature, sets the scale of the energy transfers,
and if $T\gg\Lam$ this scale is fixed to be of order $T$. The integrals
over the
four-dimensional momenta degenerate into integrals over
three-dimensional momenta.  This produces new linear infrared
divergences in high-temperature QCD \cite{kapusta}. Whether
this parallel leads to non-trivial insights into our
problem remains to be seen.

The fact that corrections linear in $1/m_Q$ are present in some
quantities is not surprising by itself. Perhaps, the best-known
example is the axial constant $f_Q$ ($f_D$ or  $f_B$ for charm and
beauty, respectively). In the
limit $m_Q\ra\infty$ they
scale  with $m_Q$ as follows \cite{A,B,VS}:
\beq
\frac{f_B}{f_D}=(\frac{m_c}{m_b})^{1/2} \;
(\frac{\alpha_s(m_b)}{\alpha_s(m_c)})^{2/b}
\label{c1}
\eeq
This scaling is known  to be significantly modified by pre-asymptotic
$1/m_Q$ terms if
one defines the axial constant in the standard way via matrix
elements
of the
axial current,
\beq
\matel{0}{\bar{Q} \gamma_\mu \gamma_5 q}{P_Q}= if_Qp_\mu
\label{c2}
\eeq
However the correction is  much smaller if one would define it via
the
pseudoscalar
current \cite{mn1,mn2}:
\beq
\matel{0}{\bar{Q} i\gamma_5 q}{P_Q}= \tilde{f}_Q M_{P_Q}
\label{c3}
\eeq
In the heavy quark limit the two definitions obviously coincide,
\beq
\frac{\tilde{f}_Q}{f_Q}\simeq \frac{M_{P_Q}}{m_Q+m_q}
\label{c4}
\eeq
and this difference is therefore indeed contained in $1/m_Q$
terms.

Calculation of $f_Q$ cannot be directly
formulated as an OPE-based procedure; therefore, the emergence of
the $1/m_Q$ correction is natural. The pole mass belongs to the same
class. Yet the OPE prescription is fully applicable to the
calculation of the inclusive widths of heavy flavor
hadrons; the
infrared
behavior of the corresponding Wilson coefficients is given by matrix
elements of the appropriate
operators. This difference between the pole mass and the
inclusive
widths is not
accidental, of course, and could be anticipated.

A remark is in order here concerning the place the pole mass
occupies in the heavy quark
effective theory \cite{HQET,HQETr}.
HQET is formulated in such a way as
to get rid
of the heavy quark mass at all; the effective theory is  left without
a large
parameter which might set an appropriate scale for distances. It is
then
only
the renormalization point $\mu$ that fixes the scale of momenta for
static
quantities. However the pole mass assumes that one takes the limit
$\mu\ra 0$
and, thus, no parameter  is left at all. It is clear that in any
consistent
formulation of HQET it is impossible to set $\mu = 0$. One should
keep $\mu$ explicitly and operate only with $m_Q(\mu )$.

It is well known that in the
framework of the HQET a key role is played by the difference
between the
the hadron and the quark mass for an asymptotically heavy quark:
\beq
\La=\lim_{m_Q\ra\infty} (M_{H_Q}-m_Q) \;\;.
\label{o1}
\eeq
It is always stated that the mass $m_Q$ entering this definition is the
pole
mass. We have shown, however, that this quantity is ill-defined.
There have been a few attempts to define it consistently in HQET.
One of
those has been made in \re{N51}; namely, for pseudoscalar mesons it
was suggested that
\beq
\La_P=\frac{\matel{0}{i(v\partial )(\bar{q} i\gamma_5 h_v)}{P(v)
}}{\matel{0}{\bar{q} i\gamma_5
h_v}{P(v)}} .
\label{o40}
\eeq
The fields of the HQET are assumed here and standard notations
are used.
This definition {\em per se} is not better and is plagued by the same
problems -- the impossibility of disentangling non-perturbative
effects from the perturbative contributions. Any consistent
formulation has to discriminate large and small momenta rather than
perturbative versus non-perturbative effects. A consistent
formulation must then
include the renormalization point $\mu$ to ensure that momenta
higher than
$\mu$ do not appear in the matrix elements.  If $\mu$
is introduced one must then define a ``running'' value of
$\La(\mu )$,
depending on the renormalization point $\mu$,  as follows
\beq
\La(\mu)=\lim_{m_Q\ra\infty} (M_{H_Q}-m_Q(\mu)) \;\;.
\label{o1a}
\eeq
with $\mu$ having to exceed sufficiently typical hadronic scales.
Its
renormalization
point dependence is given by \eq{o22}; attempts to put $\mu$ to
zero to
arrive at the `old' $\La$, would bring about
uncontrollable uncertainties of order $\Lam$ in $\La (0)$.

It has been suggested   \cite{N54} to use the requirement that
the so-called residual mass term vanishes
to rigorously define the heavy quark {\em pole}
mass $m_Q^{pole}$. In Sect. 5 it was shown that mixing between
the operators $\bar Q i\not\!\! D Q$ and $\bar Q Q$ arises
already at the one-loop level to order $\mu$.
This means that even if the effective Lagrangian of HQET is chosen in
such a way that at a certain $\mu$ the residual mass is zero, it
necessarily re-appears at a different value of $\mu$.
If the process is characterized by a single scale (like $m_Q$ in the
total
inclusive widths) one can certainly adjust the effective Lagrangian
so that there is no residual mass. If the scale varies, however, it
is mandatory to ``readjust'' the notion of the mechanical mass of the
heavy quark. In other words by requiring the vanishing of the
residual mass  one
defines the running quark mass. (It has been traced \cite{mn3} how various
observables calculated in HQET via
matching
with full QCD turn out to be independent of this term although it
is
present in intermediate stages \footnote{N.U. is grateful to
V.M. Braun
for pointing out this constructive interpretation.}.)

The fact that perturbative corrections to $m_Q$ vanish in
dimensional
regularization, which is often referred to as a remedy, is an
accident due to the fact that HQET does not contain a
dimensional parameter at the
perturbative level. References to
dimensional regularization {\em per se} without an actual subtraction
scheme is
irrelevant for the effective theory which requires a clear separation
of
high and low momenta. For example, the $\overline{MS}$ scheme
giving a
logarithmically dependent
one-loop value for $m_Q$ in the form $m_Q(\mu)=m_Q(m_Q)
(1+(2\alpha_s /\pi )\log(m_Q/\mu))$  for an infrared
cut-off $\mu\ll m_Q$
cannot achieve the momentum separation necessary for OPE.

Note that the problem of the total  inclusive widths is, strictly
speaking, outside pure HQET: for the
large
mass scale parameter -- the energy release --
is intrinsically involved
from
the very
beginning, determining the characteristic space-time separations here.

Some relations are known for exclusive formfactors in $b\ra c$
transitions, that, in the limit $m_b,m_c\,\ra\infty$ depend on $\La$
outside
the zero recoil kinematics
(see e.g. \cite{Lambdaff}). However they contain purely perturbative
${\cal O}(\alpha_s^n)$ corrections. Their main piece reflects the hybrid
renormalization \cite{VS,politzer} that must be properly accounted
for in
constructing the effective low energy description; they are governed
by the
scale momenta between $m_b$ and $m_c$, or, in general, above
$\mu$. Some
corrections, appearing already in the effective theory itself,
reflect, however, much lower scales;
obviously in the heavy quark limit
at $v\ne
v'$ there always exists a large domain of gluon momenta $0\lsim
|\vec{k}|
\lsim |\vec{v}| m_Q$ whose effect is not governed by
$\alpha_s(m_Q)$,
but rather by the running coupling at the lower scale. These
corrections
therefore can lead again to infrared
renormalon contributions. There is no reason for them not to be
able to
convert the ``pole'' $\La$ into a `running' one. It is clear that this
problem
deserves further theoretical studies, especially if one wants to
understand
the real meaning of the relations mentioned above in the presence of
an infrared
renormalon in the heavy quark mass. In principle a more
complicated situation
is conceivable -- {\em exclusive} modes may not be described
completely by the
standard OPE and intrinsic uncertainties similar to that in the
pole mass
could have emerged.

We do not address in this paper the numerical aspects of the infrared
renormalons;
they are left for the future studies \cite{irnew} (see also
\cite{BBraun}).
Still for orientation it is instructive to consider  the
estimate (\ref{linear}). According to this
expression the uncertainty in the pole mass is
\beq
\delta m_Q^{pole}\sim
\frac{8}{3b}\Lam .
\label{delm}
\eeq
To determine what particular $\Lambda$ (i.e. $\Lambda$
corresponding to what particular subtraction scheme) enters on the
right-hand side
one  needs to perform a two-loop
calculation.
In the absence of such calculations it is consistent to use
the one-loop $\Lambda$ which is one and the same in all schemes,
$\Lambda_{one\; loop}\sim $ 100 to 150 MeV. Then
one gets
\beq
\delta m_Q^{pole} \sim 50\MeV ;
\label{num}
\eeq
certainly this estimate needs improvement via real two loop
calculations.

The presence of the $1/m_Q$ renormalon in the pole mass of the
heavy quark
makes  the attempts to extract the  value
of
the pole mass from the absolute
widths of heavy flavors \cite{Ds,Calif,ligeti} not very meaningful from
a theoretical perspective; it is the masses $m_c(m_c^2)$ and
$m_b(m_b^2)$ (see
\re{Ds}) that can be extracted from the widths.
Indeed,  one can reach the required $1/m_Q$
accuracy through
calculating sufficiently many perturbative
corrections only
if the
semileptonic width of $D$ or $B$ mesons is expressed in terms of the
high
scale mass.  On the other hand, one {\em could} determine
$V_{cb}$
from the $b\ra c$ semileptonic width
assuming that one
is close to the small velocity (SV) limit \cite{SV}
 --
then this width depends mainly on the difference $m_b-m_c$
meaning   that the overall
uncertainty in the mass cancels out in $m_b-m_c$
\cite{Ds,Calif,BUV}.

The uncertainty in the pole mass can well
constitute
a sizable part of the commonly accepted value for $\La$ for mesons
of about
$400\div 600\MeV$, cf.  \eq{num}.
This is apparently due to
the peculiar Coulomb enhancement of the
leading radiative corrections that has been noted in
a different  context by Braun {\em et al.} \cite{Braun}.

\vspace*{0.5cm}

{\bf ACKNOWLEDGEMENTS:} \hspace{.4em}
During  work on this paper we were informed by V. Braun about
similar
studies undertaken by him together with M. Beneke \cite{BBraun}.
They address
similar problems from more formal positions, and arrive at similar
conclusions concerning the problem of the proper mass definition.
We are
grateful to these authors for communicating some of their results prior to
publication.
N.U. is grateful to V.
Braun  for
interesting discussions in connection with the
paper
\cite{bigi} and subsequent exchange of ideas. A.V. would like to thank V.
Zakharov for discussions.

\vspace{1mm}

This work
was supported in part by the National Science Foundation under the
grant number
PHY 92-13313 and by DOE under the grant
number DOE-AC02-83ER40105.

\vspace{3cm}

{\bf Figure Captions}

\vspace{2mm}

Fig. 1 . Feynman diagrams for $\Pi_{\mu\nu}$ in eq. (2). Solid lines
denote the quark $q$, dashed lines gluons.

\vspace{2mm}

Fig. 2 . ``Bubbles" in the gluon Green function giving rise to the
infrared renormalon in $\Pi_{\mu\nu}$.

\vspace{2mm}

Fig. 3 . One-loop diagram for the mass renormalization. Thick line
denotes the heavy quark.

\vspace{2mm}

Fig. 4 . Infrared renormalon in the pole mass.

\vspace{2mm}

Fig. 5 . Diagrams determining ${\cal O}(\alpha_s )$  corrections to the total
semileptonic width (taking the imaginary part is implied).

\vspace{2mm}

Fig. 6 . Graphs 5$a$ and $d$ in the language of the effective local
vertices (denoted by the black box).

\end{document}